# Cross Sensory Co-Design Tools and Interaction Qualities

ALBRECHT KURZE, Chemnitz University of Technology, Germany
KLAUS STEPHAN, Chemnitz University of Technology, Germany

What is the temperature of a loud cry? What is the color of a hand-clap? Cross-sensory and multi-modal interactions offer interesting possibilities in terms of inputs and outputs that might be leveraged for affective and emotional purposes. However, it is hard to design such interactions without the right co-design tools. We developed such tools in the last 10 years: the Loaded Dice and the Wheel of Plush. Both devices incorporate a multitude of sensors and actuators to demonstrate interactions and aid in co-design workshops. The tools incorporate the principle of technical synesthesia: the ability to map any of the included sensors to any of the included actuators. With intuitive simple interaction schemes it is possible to easily and effortlessly create new combinations. We discuss the core principles of the tools and how we realized a meaningful mapping between sensors and actuators. We further discuss how we use the tools for exploration, sensory sketching or scenario driven ideation, and where we see additional potential.

## 1 Introduction

The "traditional five" human senses are sight, hearing, taste, smell and touch. Secondary senses are temperature, pain, proprioception and balance [19]. Synesthesia describes the phenomenon of an event being experienced by another, separate sensory modality [5]. While medically not exact, in principle, this means a sound might not only be heard but also be seen as a color (as an example). Most existing design tools, i.e. for sensory ideation, do not employ synesthesia effects as a design opportunity in order to break with existing sensing stereotypes for framing design spaces. Such a stereotype could be e.g. that making noise should always be connected with hearing noise. While most related tools, e.g. *littleBits* [2], do allow for flexible combinations of different sensors and actuators in principle, it is not possible ad hoc. Instead they require additional steps in combining parts or mapping sensor values to actuator values.

This was our starting point ten years ago when we designed and developed the *Loaded Dice* [14–16], a multi-sensory and multi-modal hybrid toolkit to ideate multi-sensory and cross-modal devices and scenarios, e.g. for 'smart' devices, and with different groups of co-designers [3, 4, 12, 15]. The *Loaded Dice* filled a gap between analog, non-functional tools, often card-based [13], e.g. *KnowCards* [1], and functional but tinkering based tools for multi-sensory und multi-modal exploration, ideation and prototyping, e.g. the before mentioned *littleBits*. Over the years we continued our work on the Loaded Dice, conceptualized new embedded interaction qualities further, optimized the mapping between inputs and outputs [10] and experimented with the addition of different material qualities, i.e. with textiles and here in specific with plush. In a follow-up project we picked up the core idea of a flexible co-design tool and extended it even further in the "Wheel of Plush" [20]. It now offers 8 inputs and 8 outputs in a variety of modalities, integrates them all in plush

Authors' Contact Information: Albrecht Kurze, Chemnitz University of Technology, Germany, Chemnitz, albrecht.kurze@informatik.tu-chemnitz.de; Klaus Stephan, Chemnitz University of Technology, Germany, Chemnitz, klaus.stephan@informatik.tu-chemnitz.de.

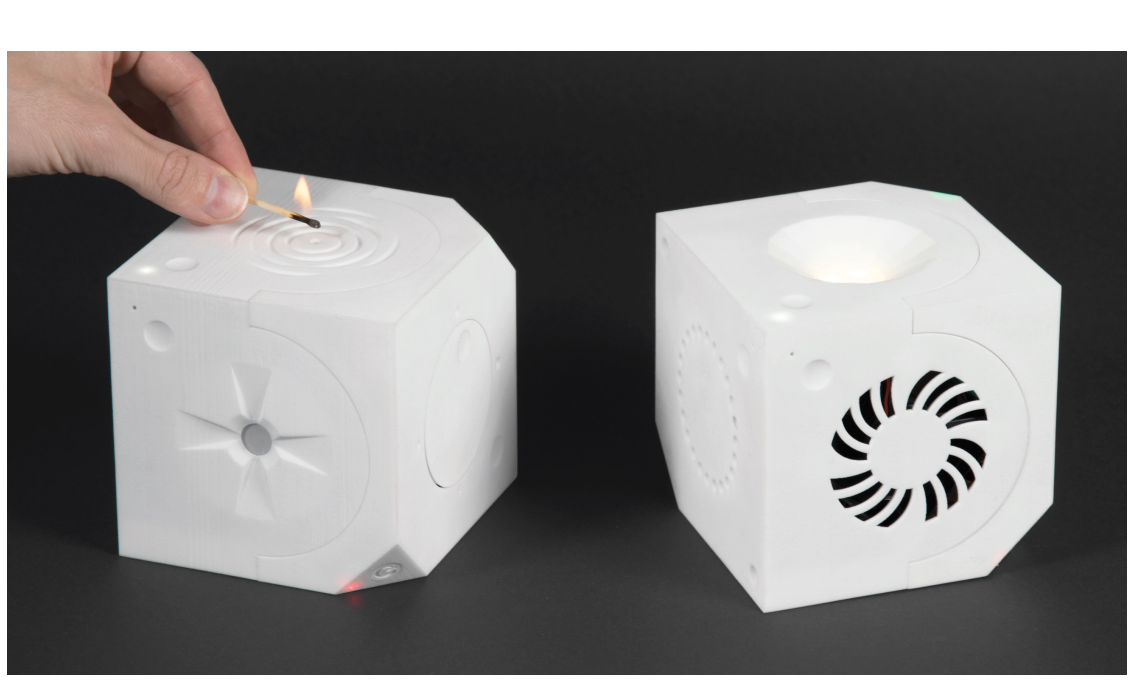
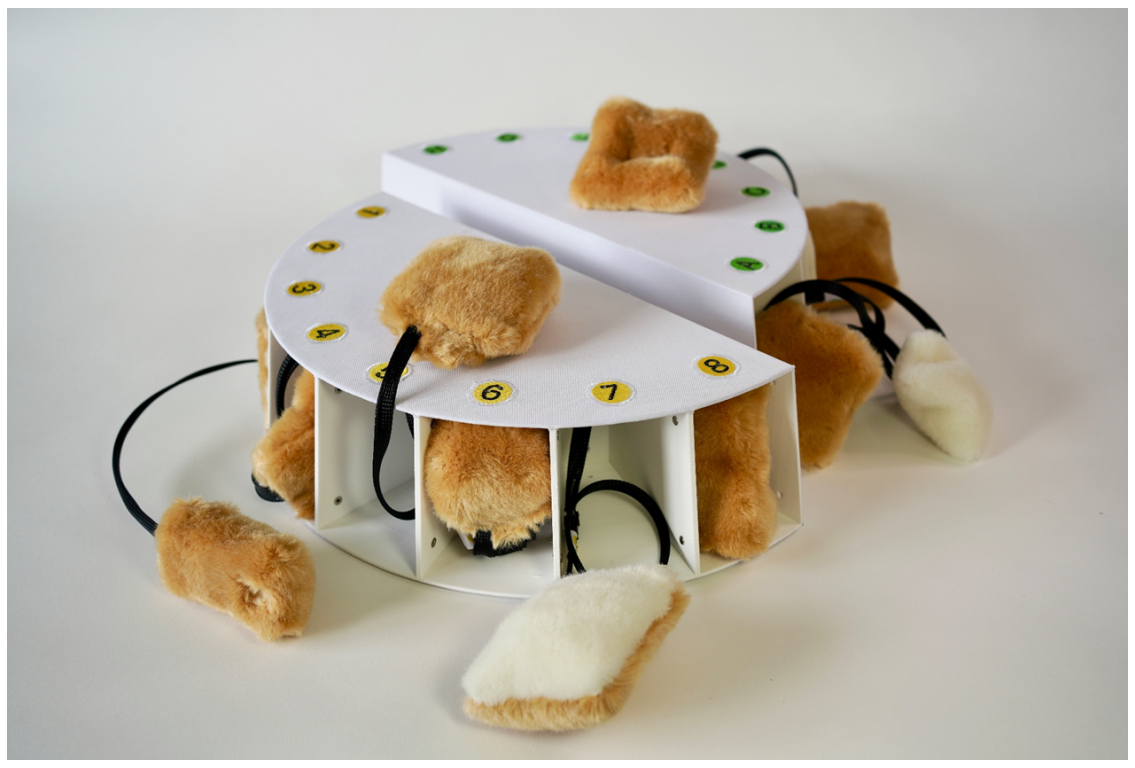

Fig. 1. left: The Loaded Dice, example of devices in use, turning heat into light (sensor die with temperature sensor active and actuator die with power LED active) [15]; right: The Wheel of Plush, consisting of two semi-circular sections, each with 8 compartments for 8 sensor and 8 actuators respectively [20]

for a specific material experience, allows multidimensional normalized mappings from inputs to outputs [21] as well as a flexible combination of multiple outputs to create not only cross-sensory but also multi-modal experiences.

In a series of co-design workshops, we found preferences in inputs, outputs, modalities, and their combinations, which proves the value of these tools.

This brings us to our core question that we followed: *How can we build on these devices to design even better cross sensory experiences, and support or supplement innovative interactions and non-verbal communication?*

In this submission we briefly summarize the Loaded Dice and the Wheel of Plush, the core concepts that they are built on, the used sensors and actuators, and how they map to different human senses. We will then continue to discuss how these tools can be used to design cross-sensory interactions.

## 2 The Loaded Dice

The *Loaded Dice* [1] are two cubical devices which are wirelessly connected. Each cube has six sides, offering six sensors in one cube and six actuators in the other cube which address different human senses (Table 1), one on each side, suitable for multi-sensory and multi-modal environmental and user interactions. The cubical shape communicates the intuitive reading that the top side is active, like a die, offering an easy and spontaneous way to re-combine sensors and actuators by just turning a new face up. Every sensor-in and actuator-out combination is possible in an 1-on-1 combination resulting in 36 combinations in total [15]. A combination of multiple inputs or multiple outputs is technically possible, but we explicitly decided not to implement them due to interaction and shape constrains (e.g. side facing away from a user or even downwards to the desk etc.).

The sensor cube normalizes a raw sensor value meaningfully, transmits it, and then the other cube actuates it mapped on an output. We implemented a meaningful mapping between every sensor and actuator that is used in the Loaded Dice. This includes reasonably chosen sampling rates, ranges and steps for raw input values, their normalization on internal values and the conversion back to meaningful output values. All this is done internally in hard- and software, without the need of user intervention. Selecting a new sensor-actuator combination just requires bringing another side

[1]video demonstrating the Loaded Dice: https://www.youtube.com/watch?v=-E5aUiktCic

Table 1. Human senses vs. sensors and actuators in the Loaded Dice [10]

| Human Sense | Sensor | Actuator |
|---|---|---|
| sight<br>(visual stimuli) | luxmeter (visible light luminosity/brightness)<br>passive infrared detector (PIR movement)<br>ultrasonic transceiver (distance) | power LED (brightness)<br>LED ring-graph (count, overall brightness, color) |
| hearing<br>(auditive stimuli) | microphone (amplitude / loudness) | sound (modulated note for instrument)<br>(vibration motor, rattling noise)<br>(fan, air flow noise) |
| touch<br>(tactile stimuli) | potentiometer (manual angular dial of 270°) | vibration motor (vibration)<br>fan (mechanical stimulation on hairs) |
| temperature<br>(thermal stimuli) | thermometer + infrared thermometer<br>(thermopile for thermal radiation) | Peltier element (cooling and heating plate)<br>fan (cooling by chill effect on skin) |

to the top. Based on the presented design rationale, a co-designer can transport heat over a distance by choosing the infrared thermometer and Peltier element sides of both cubes. Rotating the actuator cube to the power-LEDs would transform the temperature into light, thus mimicking synesthesia-like perception.

For the Loaded Dice we decided deliberately for a one-dimensional mapping with only 25 steps but adjusted to nonlinearities in human perception (e.g. for light) and addressing technical limitations (e.g. minimum motor speeds for mechanical actuation/vibration). Details about the mapping are published in [10].

In 2025 we updated the technology in the Loaded Dice and simplified their build process, reducing their bill of materials and costs of production significantly to prepare for open sourcing them.

## 3 The Wheel of Plush

Three years ago, we started developing the "Wheel of Plush" (WoP), our new co-design tool for the design space of smart and connected plush toys [11]. We decided early on to build on the topology of the Loaded Dice [16] with an input/sensor side and an output/actuator side and both sides communicating wirelessly. The Wheel of Plush features 8 sensors in one part and 8 actuators in the other part. Each sensor and actuator is ‘plushified’, which means it is built into a plush cushion or another plush-covered form to explore what technologies can be utilized with this challenging material. We explored the application of textile technologies in plush, which due to its fluffy nature makes many techniques much harder (obstructs sensors, adds thick padding, easily damaged by heat, adhesives, sowing or embroidering). An in-depth explanation with additional pictures can be found in our pictorial [20].

Learning from the Loaded Dice we conceptualized the Wheel of Plush to build a foundation with some new traits for cross-sensory design:

**New and more sensors and actuators:** The Wheel of Plush comes with a number of new sensors (e.g. an internal inertial measurement unit (IMU), force and bending sensors) as well as new actuator (servo and pneumatics) [22, 23]) that open up new possibilities for interaction. Despite these do not addressing completely new senses or modalities, they surpass the abilities of the Loaded Dice.

**Modularity**: The individual sensors and actuators are easily exchangeable and swappable using 4 pin connectors and a set of defined interfaces (basic digital, basic analog, serial, I2C). This allows to swap a specific sensor or actuator to another one with a different material appearance (other plush, other textiles, other construction) but it also enables some extensibility as long as the interface is usable. (This is the case for many standard sensors using I2c.)

Table 2. Sensors and actuator in the Wheel of Plush

| Sensors | Human Sense Equivalent | Actuators | Addressed Modalities |
|---|---|---|---|
| 1) Distance / Light Sensor | sight | A) Sound | audible |
| 2) Temperature Sensor | temperature | B) Single LED | visible |
| 3) Flex Sensor | touch + proprioception (force) | C) LED Array | visible |
| 4) Pressure Sensor | proprioception (force) | D) Vibration | tactile feelable, not audible |
| 5) Touch Sensor | touch | E) Heating | thermo perceptable |
| 6) Tactile Sensor (binary) | proprioception (force) | F) Fan | tactile feelable, slightly thermo perceptable, slightly audible |
| 7) Motion Sensor | balance | G) Servo | visible, tactile, audible during changes |
| 8) Sound Sensor | hearing | H) Pneumatic | tactile, visible, slightly audible during pumping and draining |

**Multidimensional data normalization and mapping:** We also decided to keep but extend the concept of software-defined logical coupling on the data level instead of hardwired coupling. While the Loaded Dice use only an one-dimensional mapping we implemented in the Wheel of Plush a multi-dimensional mapping [21]. For some sensors we still use only one dimension, for other two (e.g. bending in x- and y-direction) or even three (e.g. for the IMU with x,y,z-axis). The same time we made the internal value space much more granular (65536 distinct value instead of only 25). This does not mean we really have to differentiate between the same number of steps in the input or output (there is often a much lower number sufficient) but to allow complex, non-linear mapping functions with high accuracy in completely different shapes. We also introduced an intermediary device in the communication between the sensor side and actuator side based on a Raspberry Pi running Node-Red. This allows to introduce and customize new mapping functions on the fly – either in simple flow-driven programming from a source (sensor) to a sink (actuator) using existing building blocks or simply through custom JavaScript code snippets.

**1:n mapping:** Similar to the Loaded Dice the sensors and actuators can be paired up in any of the 64 possible 1:1 combinations. However, this is only one operational mode (default). Soon after a few first use in a workshop the wish emerged to allow other combinations than only 1:1, especially to combine multiple outputs. We implemented a new actuator selection scheme for this purpose. A short touch on a selection element changes the selected actuator, e.g. from A to B. A long press 'freezes' this selection and allows to add further actuators, e.g. A+B+D, until another long press 'unfreezes' a selection. This allows, e.g. to combine vibration in form of a pulsation with colorful light and a sound modulation – all driven by a single sensor. Such combinations allow us to address multiple senses in parallel what can help to emphasize or modulate selected experiences in contrast to a 1:1 mapping. Together with the flexible "sketching" of a new mapping function this would allow to bring in completely new synesthesia effects, also not necessarily only limited to one input sensor and one output actuator at one time.

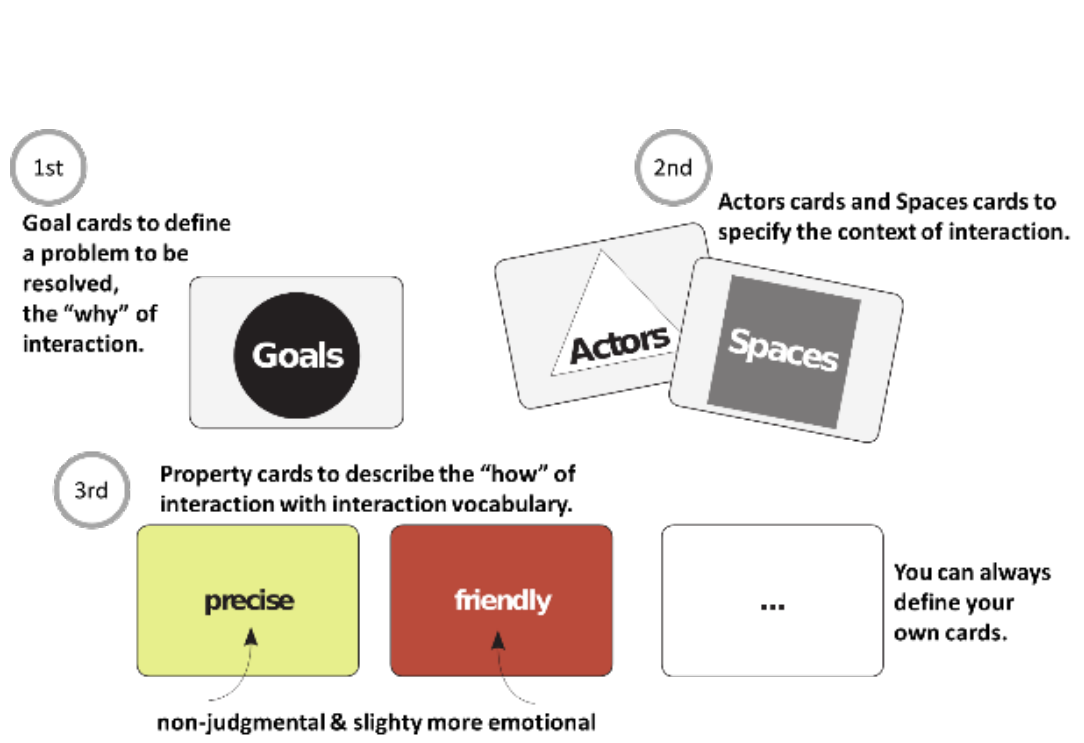

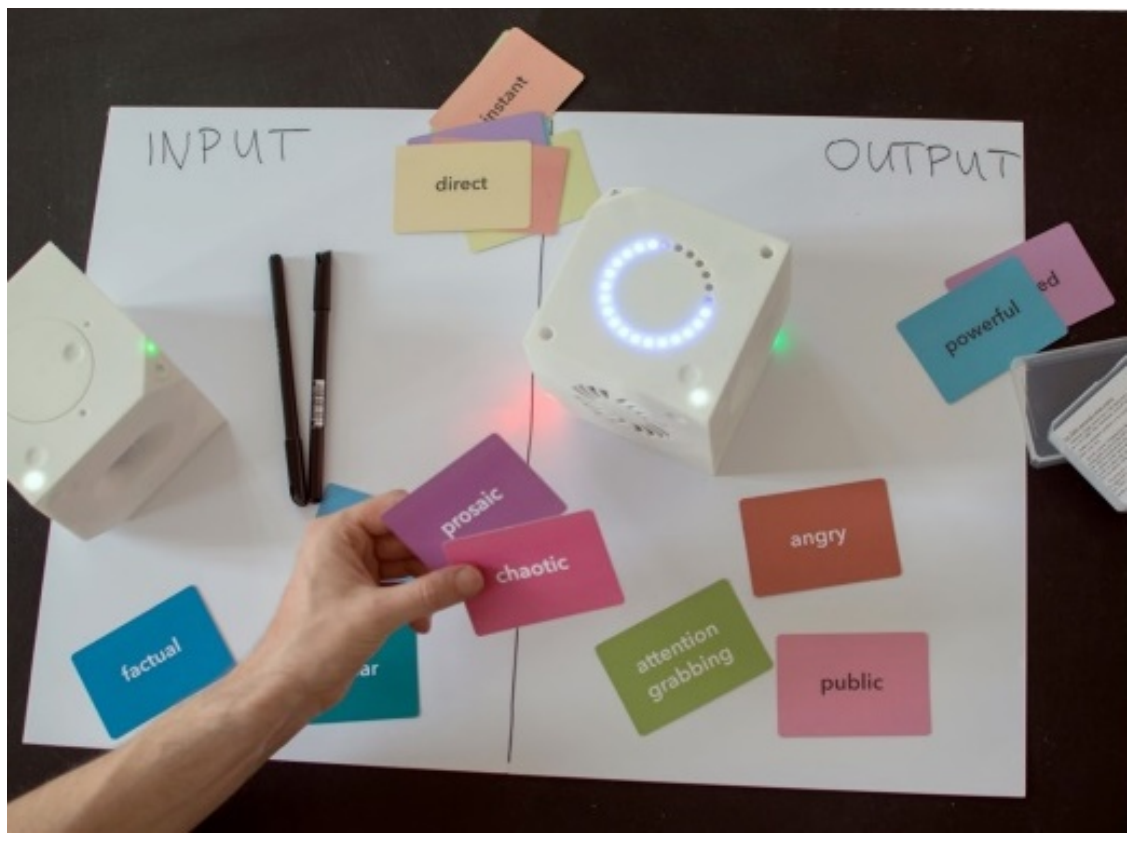

Fig. 2. left: the types of cards of our co-design method, 1st setting a goal (why), 2nd context of interaction (actors and space), 3rd defining desired interaction qualities; right: cards with terms of the extended interaction vocabulary [8], an example of using the cards to characterize and map input and output qualities of an ideated connected product with help of the Loaded Dice [9]

## 4 Cross-Sensory Design and Interaction Qualities

Over the years we used these both tools in a series of co-design workshops but also in teaching settings. This way we learned a lot of the potential uses in these settings and also what modes of operation work best in different settings and with different target groups and stakeholders. The possibilities of the Loaded Dice and the Wheel of Plush can be used in different co-design settings (WoP = Wheel of Plush only):

a) in open exploration, mainly to explore how (well) something (might) work
   - to try out sensor or actuator functions
   - determining technical limitations
   - finding possible usages
b) for sensory sketching, creating ad hoc mappings, even for 'weird' synesthetic combinations, e.g:
   - to try out how bright sunlight sounds or feels as vibration
   - what temperature a loud cry has
   - how much air-flow half a meter distance is
   - whether you can feel in vibration how the lights flicker
   - what pulse rate a hard bend is in contrast to a soft one (WoP)
   - what consistency a quick movement is (WoP)
   - what color a clap has (WoP), ...
c) in a framed scenario-driven co-design approach, that integrates a card-based component to structure the design process and includes an adapted interaction vocabulary and in a final step an adjustment of sensory modalities to desired interaction qualities (see Figure 2).

We used these different modes successfully in co-design workshops, also in combination. This brought a number of different findings: which sensors or actuators are preferred or avoided for certain products or scenarios, which scenarios people ideate at all in the interaction with, between and through smart devices [8], what combinations are typically associated with specific interaction qualities, what interaction modalities and qualities are associated with which emotion [9] and how it is possible that co-designers are immersed in the overall process.

## 5 Conclusion and Outlook

We presented and discussed the Loaded Dice and the Wheel of Plush as cross/multi-sensory/modal co-design tools. Together with the associated methods and interaction quality mappings they proved as value tools to explore the design space of new cross-sensory interfaces and interactions.

New and other modalities (smell, taste) are still hard to achieve. At least in terms olfactory inputs there are significant advances in the recent generation of metal oxide semiconductors (MOS) sensors that can already work as simple "e-noses", and thanks to embedding of AI also allow for a bit more complex uses [18]. New smaller piezo evaporators allow to emit scents in small spaces on the actuator side (but the problem of active neutralization or sufficient ambient air exchange persists). Nevertheless, these modalities have the potential to broaden interaction qualities even further, especially in an emotional way [7, 9, 19] as well as cross-modal between smell and shape [6].

Despite of its possibilities we consider the Wheel of Plush as co-design tool that best fits the design space of smart textiles and soft toys that we tailored it for [11]. All the advantages that we introduced come at a cost: higher technical complexity, higher costs, much higher fabrication demands, less compact, less transportable.

Nevertheless, we think it is time to pick up all strong points of both concepts and to combine them in a next generation co-design tool: compact, integrated design of the Loaded Dice, interaction simplicity of the Loaded Dice combined with n-up parallel output modality selection of the Wheel of Plush, modularity in terms of swappable sensors and actuators for a greater multitude of sensors and actuators, and multi-dimensional normalized mappings

Some of these aspects are already achievable using just more than one pair of the Loaded Dice. Other aspects might require a redesign that introduces swappable faces, combination of smaller cubes as modules in a Borg-style fusion cube or rearrangeable sub-cubes, e.g., with joints that can be rotated around different axes, similar to Rubik's cubes - an advantage of the cubical base shape [17].

## Acknowledgments

This research is funded by the German Ministry of Research, Technology and Space (BMFTR) grant FKZ 16SV9117.